\documentclass[twocolumn,twocolappendix,trackchanges]{aastex631}

\usepackage{graphicx}	
\usepackage{amsmath}	
\usepackage{physics}
\usepackage{gensymb}
\usepackage{isotope}
\usepackage{color}
\usepackage{cleveref}
\usepackage{multirow}
\usepackage{comment}

\newcommand{\x}{\times}
\newcommand{\gsim}{\gtrsim}
\newcommand{\lsim}{\lesssim}

\newcommand{\feh}{\rm[Fe/H]}

\newcommand{\mstar}{m}
\newcommand{\deltat}{\Delta t}

\newcommand{\Msun}{\,{M_{\odot}}}
\newcommand{\Lsun}{\,{L_{\odot}}}
\newcommand{\au}{\,{\rm au}}
\newcommand{\parsec}{\,{\rm pc}}
\newcommand{\kilopc}{\,{\rm kpc}}
\newcommand{\megapc}{\,{\rm Mpc}}
\newcommand{\yr}{\,{\rm yr}}
\newcommand{\gigayr}{\,{\rm Gyr}}
\newcommand{\megayr}{\,{\rm Myr}}
\newcommand{\erg}{\,{\rm erg}}

\begin{document}

\title{Are Local Group Dwarf Spheroidal Galaxies the First Safe Planet-hosting Environments?}

\author[0009-0000-3803-786X]{Stefano Ciabattini}
\affiliation{Dipartimento di Fisica e Astrofisica, Università degli Studi di Firenze, Via G. Sansone 1, 50019 Sesto Fiorentino, Italy\\}
\affiliation{INAF/Osservatorio Astrofisico di Arcetri, Largo E. Fermi 5, 50125 Firenze, Italy\\}
\email{stefano.ciabattini@unifi.it}

\author[0000-0001-7298-2478]{Stefania Salvadori}
\affiliation{Dipartimento di Fisica e Astrofisica, Università degli Studi di Firenze, Via G. Sansone 1, 50019 Sesto Fiorentino, Italy\\}
\affiliation{INAF/Osservatorio Astrofisico di Arcetri, Largo E. Fermi 5, 50125 Firenze, Italy\\}

\author[0000-0003-1859-3070]{Leonardo Testi}
\affiliation{Alma Mater Studiorum, Università di Bologna, Dipartimento di Fisica e Astronomia (DIFA), Via Gobetti 93/2, 40129 Bologna, Italy \\}
\affiliation{INAF/Osservatorio Astrofisico di Arcetri, Largo E. Fermi 5, 50125 Firenze, Italy\\}



\begin{abstract}

We explore whether Local Group dwarf spheroidal (dSph) galaxies might have hosted Earth-like planets dwelling unexposed for several billions of years to major galactic threats to life, such as supernovae and gamma-ray bursts. To this aim, we developed a novel semiempirical model that exploits the observed chemical abundances and star formation histories of a selected sample of local dSphs, to explore whether their stars may have (i) reached the minimum metallicity to trigger planet formation and (ii) avoided exposure to destructive events long enough to provide time for possible biological development. From our work two scenarios emerge. If planet formation is possible for ${\rm[Fe/H]}\lesssim-1$, then in all dSphs with $5\x10^{3}L_{\odot}\leq L_V\leq2\x10^{7}L_{\odot}$ a fraction $\approx0.1\%-10\%$ of stars might have safely hosted terrestrial planets for more than $1$ Gyr. In this scenario, ancient ultra-faint dwarf galaxies (UFDs, $L_V\leq10^{5}L_{\odot}$) would have been the first to reach this condition in the history of the Local Group. Conversely, if planets form for ${\rm[Fe/H]}\geq-0.6$ then they should not exist in UFDs, while only $\approx0.001\%-0.1\%$ of stars in dSphs with $L_V\geq3\x10^{5}L_{\odot}$ would host planets dwelling in safe conditions for long times. Interestingly, we find a "luminosity sweet spot" at $L_V\sim10^{6}L_{\odot}$ where dSphs in our sample safely host terrestrial planets up to $4$ Gyr and in any planet formation scenario explored. In conclusion, planet formation at low metallicity is key to understanding which types of galaxies might have formed Earth-like planets that dwelt unexposed to galactic threats over several billions of years, first in the history of the Local Group.

\end{abstract}



\section{Introduction} \label{sec:intro}

Habitability is generally associated with the stellar habitable zone, i.e. the region around a star where liquid water can exist on the surface of a terrestrial (i.e. Earth-like) planet for an extended period of time \citep[e.g.][]{Huang59,Kasting+93,Kopparapu+13}.
Hence, such concept is linked to the scale of a single-planetary system.
However, \citet{Gonzalez+01} for the first time considered habitability on a larger scale, extending the idea to the entire Milky Way (MW). 
In particular, they identified a region of the Galaxy that is sufficiently enhanced in heavy elements to form terrestrial planets, which could provide a long-term habitat for life.
For this reason, the minimum metallicity to trigger planet formation is considered a key quantity to determine the probability of potentially habitable planetary systems to form in a galaxy.

Still, this is just one of the two necessary (but not sufficient) conditions to potentially form a life-suitable planet. 
Indeed, catastrophic events occurring during the evolution of a galaxy, such as supernova (SN) explosions, gamma-ray bursts (GRBs), accreting black holes, and merging of compact objects, might represent a serious risk to life on a planet \citep[e.g.,][]{Ruderman74,Tucker81,Gehrels+03,MelottThomas11}.
These events, in fact, inject a large amount of energy into their surroundings and could have a severe impact on the habitability of nearby Earth-like planets.
Based on these arguments, different studies have used galactic chemical evolution models to estimate the probability of habitable planets to form and evolve safely from life-harmful events, in our own Galaxy \citep[e.g.,][]{Lineweaver+04,Prantzos08,Spitoni+14,Spitoni+17,Spinelli+21} or even in other massive galaxies \citep[e.g.,][]{Spitoni+14,Dayal+15}.

However, the standard cosmological model predicts that massive galaxies such as the MW are built through the assembling of low-mass progenitor dwarf galaxies, formed at earlier cosmic epochs \citep[e.g.,][]{MoWhite02, Helmi08, Salvadori+15}. 
Some of these ancient dwarf galaxies are not incorporated into the main central object and evolve through cosmic time as satellites. In the Local Group (LG) there are $\sim100$ known satellite dwarf galaxies~\citep{Pace24}. 
Among them, the so-called dwarf spheroidal (dSph) galaxies host $>10\gigayr$ ancient and metal-poor stellar populations, and are currently lacking gas and star formation~\citep[e.g.,][]{Tolstoy+09}. DSph galaxies span several orders of magnitude in luminosity and stellar mass: from ultra-faint dSphs \citep[UFDs, $M_\star\approx10^{3-5}\Msun$, e.g.,][]{Simon19} to brighter ``classical" dSphs \citep[$M_\star\approx10^{5-7}\Msun$, e.g.,][]{McConnachie12}. 

In this work we explore whether and when LG dSph galaxies might have formed stars equipped with terrestrial planets that remained safe from destructive events during galaxy evolution.
In particular, we aim at addressing the following questions: Did LG dSph galaxies develop the conditions to potentially host life in the very early Universe? And if so, when did UFDs and ``classical" dSphs meet these conditions?
To address these issues we develop a novel semiempirical model which exploits the large amount of public-available data, for both the star-formation histories (SFHs) and stellar chemical abundances of LG dSphs. 

\section{The Data Sample} \label{sec:data}
To estimate the probability that a galaxy can safely host terrestrial planets we need to know both the metallicity distribution function (MDF) and the SFH of its stellar populations. Therefore, our dSph sample includes 11 galaxies for which these two pieces of information are simultaneously available from the literature. 
In particular, we have six UFDs and five classical dSphs, whose names and observed properties are reported in \Cref{table}.
Note that the dSphs in our sample span $\approx4$ orders of magnitude in total luminosity ($L_V$).
We also report the half-light radius of the dSphs, $R_{1/2}$, which is estimated as the scale length of the exponential profile used by~\citet{Munoz+18} to fit the observed surface brightness of the galaxy. Note that $R_{1/2}$ tends to increase for brighter dSphs. 
Finally, for each dSph we report the average iron abundance of the stars, $\langle{\rm[Fe/H]}_*\rangle$, and the number of stellar ${\rm[Fe/H]}$ measurements, $N_*^{\rm obs}$. 
\\
\\
\begin{table}
	\centering
	\caption{Observed properties of dSph galaxies from~\cite{Munoz+18} and the SAGA database.}
    \small
    \tabcolsep=0.08cm
	\begin{tabular}{l|cccc}
	\hline
	Galaxy & $\log\pqty{\frac{L_V}{\Lsun}}$ & $R_{1/2}$ & $\langle{\rm[Fe/H]}_*\rangle$ & $N_*^{\rm obs}$ \\
	   & & $\pqty{\parsec}$ & & \\
	\hline
    Coma Berenices (Com)       & $3.682$ & $72.6$ & $-2.49$ & $9$    \\
	Leo~IV                     & $3.930$ & $117$  & $-2.35$ & $5$    \\
	Ursa Major~I (UMa I)       & $3.981$ & $234$  & $-1.98$ & $17$   \\
	Canes Venatici~II (CVn II) & $4.002$ & $70.7$ & $-2.07$ & $8$    \\
	Hercules (Her)             & $4.266$ & $224$  & $-2.21$ & $26$   \\
	Bo\"{o}tes~I (Boo I)       & $4.338$ & $202$  & $-2.47$ & $76$   \\
    Draco (Dra)                & $5.417$ & $212$  & $-1.88$ & $340$  \\
	Ursa Min (UMi)             & $5.546$ & $404$  & $-1.97$ & $224$  \\
	Carina (Car)               & $5.706$ & $311$  & $-1.41$ & $1027$ \\
	Sculptor (Scl)             & $6.262$ & $311$  & $-1.80$ & $608$  \\
	Fornax (For)               & $7.317$ & $791$  & $-1.06$ & $1608$ \\
	\hline
	\end{tabular}
 \label{table}
\end{table} 

\subsection{MDFs}\label{sec:data_mdf}
\begin{figure*}
    \centering
    \includegraphics[width=0.7\linewidth]{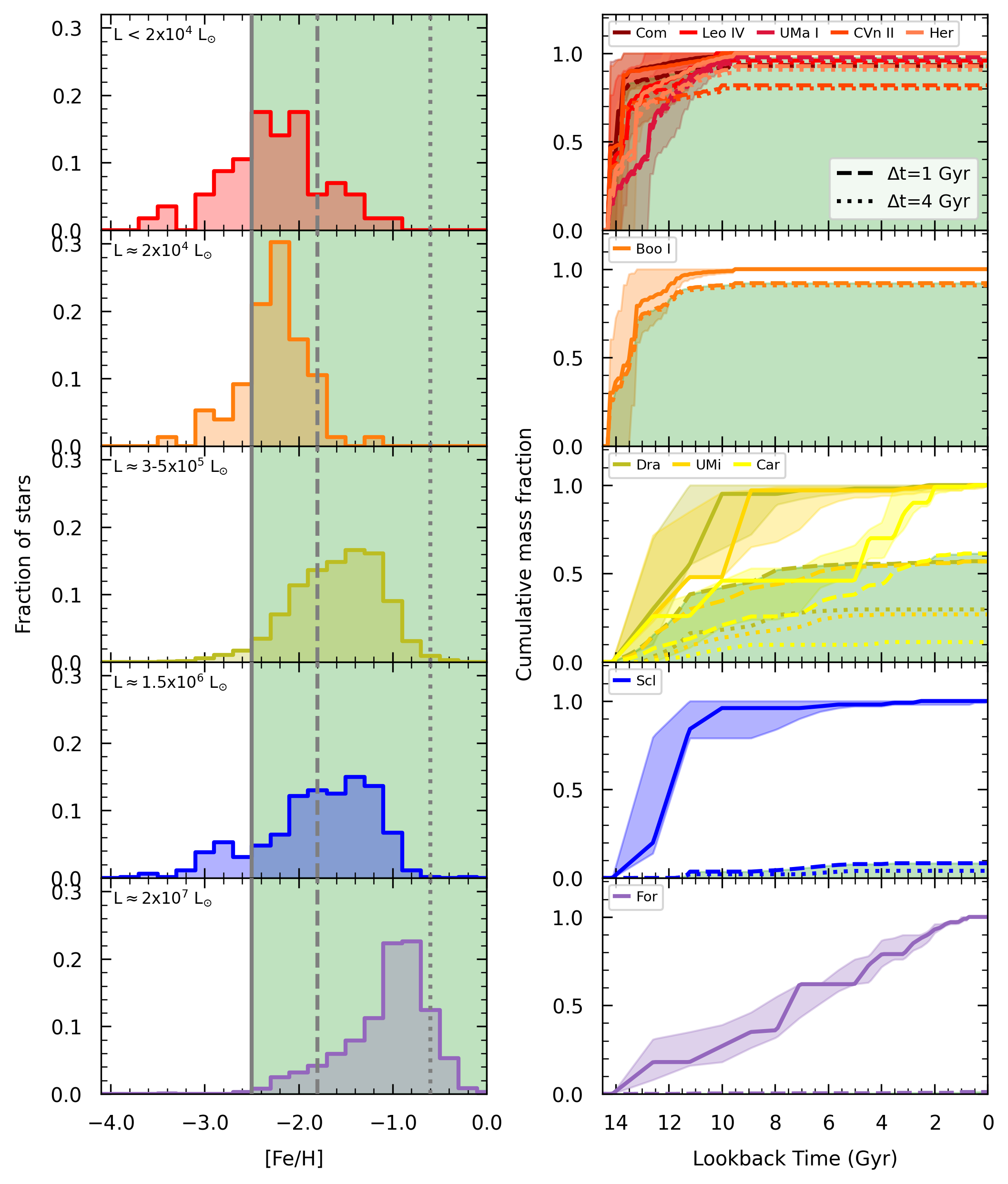}
    \caption{The MDFs (left) and the SFHs (right) of dSphs with increasing luminosity from top (red) to bottom (violet-see also labels). Left panels: all ${\rm[Fe/H]}$ measurements of stars in dSphs are taken from the SAGA database (\url{http://sagadatabase.jp/}) and have been corrected for 1D NLTE effects with the \texttt{NLiTE} tool \citep[][]{Koutsouridou25}. Vertical lines represent the minimum ${\rm[Fe/H]}$ for planet formation in models F1 and E1 (solid), F2 (dashed), and F3 and E2 (dotted; see also Sec.~\ref{sec:plformation}). The green shaded areas highlight the ${\rm[Fe/H]}\geq-2.5$ regime. Right panels: as a function of time we show the cumulative mass fraction of stars that (i) are formed in each dSph (solid lines with $1\sigma$ uncertainties, data from~\citealp{Brown+14} and \citealp{Weisz+14}) and (ii) evolve in nonsterilized regions for $1\gigayr$ (dashed) and $4\gigayr$ (dotted)-both underlined with green shaded areas.}
    \label{fig:sfh+mdf}
\end{figure*}

The MDFs of nearby dSph galaxies are obtained by measuring the iron abundance of their individual stars through spectroscopic observations \citep[e.g.,][]{Kirby+08}. 
Here we aim to build MDFs using data as consistent as possible; thus, we retrieved the ${\rm[Fe/H]}$ measurements for stars in dSphs in our sample via the SAGA database~\citep{SAGA}. 
After selecting the most recent work for stars with multiple measurements, we ended up with a sample of 3965 stars with measured ${\rm[Fe/H]}$ in total.
We corrected these measurements for the effects of 1D non-local thermodynamical equilibrium (NLTE), using the online tool {\tt NLiTE} \citep{Koutsouridou25}. For each dSph the corrected average iron abundance is reported in \Cref{table}.
The MDFs of UFDs and classical dSphs obtained from this sample of measurements are shown in the left panels of Fig.~\ref{fig:sfh+mdf}.
We can see that, despite being dominated by very metal-poor stars, UFDs feature an extremely broad MDF, which extends up to $\feh\sim-1$.
For more luminous dSphs the peak of the MDF is located at increasing ${\rm[Fe/H]}$. 

\subsection{SFHs}\label{sec:data_sfh}
SFHs of nearby dSph galaxies are obtained from their observed color-magnitude diagrams combined with the chemical properties of their stars~\citep[e.g.,][]{Tolstoy+09}. 
In order to have a sample of SFHs as consistent as possible, we used those from~\citet{Brown+14} for UFDs, and those from~\citet{Weisz+14} for classical dSphs.
Both works obtained SFHs for individual galaxies by simultaneously fitting the color-magnitude diagrams from Hubble Space Telescope observations and the stellar metallicity with stellar population models. 
Furthermore, both provided the SFHs as the cumulative stellar mass fraction formed in each galaxy versus the stellar age, which we converted here into lookback time. 

In the right panels of Fig.~\ref{fig:sfh+mdf} we show the cumulative mass fraction of stars formed in dSphs at different cosmic times \citep[][; solid curves]{Brown+14,Weisz+14}.
We see that all UFDs show a similar evolution: they form the bulk of their stars at early times and then are rapidly quenched, $>75\%$ of present-day stars being $\gsim13\gigayr$ old \citep[see also][]{Brown+14}. 
On the other hand, more luminous dSphs have longer and diversified SFHs, some with a major phase of star formation followed by minor activity until a few Gyr ago (e.g., Dra, UMi, and Scl), others with alternating epochs of star formation and quiescence (e.g. Car). 
Hence, these classical dSphs host stellar populations with a variety of ages, all of them also including $\approx13\gigayr$ old stars.

\section{The Model} \label{sec:methods}
In our semianalitical model, we compute the probability that LG dSph galaxies might have hosted terrestrial planets dwelling in a safe environment as
\begin{equation}
    P_{\rm host}=P_{\rm form}\x P_{\rm surv},
    \label{eq:Phost}
\end{equation}
where $P_{\rm form}$ is the probability of dSphs to form terrestrial planets, and $P_{\rm surv}$ the probability of planets in dSphs to survive life-harmful galactic events for a time $\deltat$. 
These two probabilities are respectively defined by exploiting the observed MDFs and SFHs of dSph galaxies (see Sects.~\ref{sec:plformation} and ~\ref{sec:ster-free}). 
Note that our model focuses on dSphs because we want to understand whether these ancient systems might have been the first to host planets in such conditions. 
However, our model is general and thus potentially applicable to any system with available MDF and SFH.

\subsection{Formation of terrestrial planets}\label{sec:plformation}
We evaluate the probability of terrestrial planets formation around stars in local dSphs as
\begin{equation}
    P_{\rm form}=\dfrac{1}{N_{\rm *,tot}}\sum_{i}f_{\rm TP}({\rm[Fe/H]})N_{*}({\rm[Fe/H]}_{i})
    \label{eq:Pform}
\end{equation}
where $N_{*}\pqty{\rm[Fe/H]_{i}}$ is the number of stars in the {\it i}th ${\rm[Fe/H]}$ bin of the MDF, $N_{\rm *,tot}$ is the total number of stars with measured ${\rm[Fe/H]}$ in the dSph, and $f_{\rm TP}({\rm[Fe/H]})$ is the probability of terrestrial planets formation as a function of the stellar ${\rm[Fe/H]}$.

\begin{figure}
    \centering
    \includegraphics[width=\linewidth]{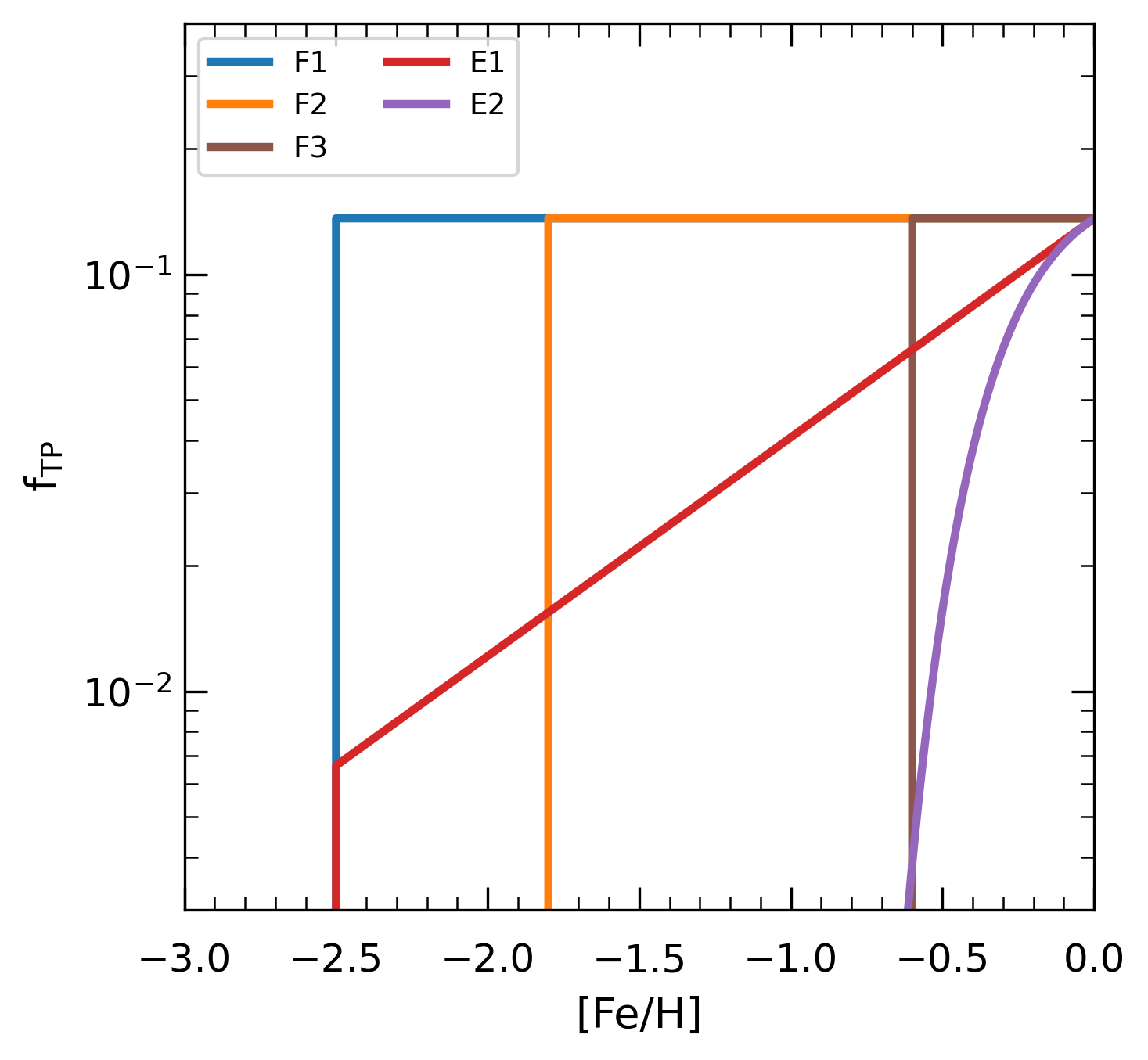}
    \caption{Adopted metallicity-dependent probabilities of terrestrial planets formation. See labels and text for the different models assumed.}
    \label{fig:eff_pl}
\end{figure}

Theoretical predictions for the minimum metallicity to trigger planet formation vary significantly in the literature.
In this work we start by assuming the simplest shape for $f_{\rm TP}({\rm[Fe/H]})$, i.e. a flat distribution above a minimum ${\rm[Fe/H]}$ (F\# models) as shown in Fig.~\ref{fig:eff_pl}:
\begin{enumerate}
    \item ${\rm[Fe/H]}\geq -2.5$ \citep[][model F1]{JohnsonLi12}, 
    \item ${\rm[Fe/H]}\geq -1.8$ \citep[][F2]{HasegawaHirashita14},
    \item ${\rm[Fe/H]}\geq -0.6$ \citep[][F3]{Andama+24}.
\end{enumerate}
Moreover, we explore two additional models that account for the metallicity-dependent occurrence rate of small-period ($<10$ days) terrestrial planets, recently observed around FGK stars in the MW \citep{Zink+23, Boley+24}. 
The observed trends have been extrapolated to lower ${\rm[Fe/H]}$ (E\# models) and they are shown in Fig.~\ref{fig:eff_pl}. We see that E\# models decline with decreasing ${\rm[Fe/H]}$ by following:
\begin{enumerate}
    \item a power law \citep[][E1]{Zink+23},
    \item an exponential law \citep[][E2]{Boley+24}.
\end{enumerate}
In Fig.~\ref{fig:eff_pl} the probabilities of all these models for terrestrial planets formation are reported as a function of the iron abundance of the hosting star and are normalized to reproduce the observed occurrence rate of terrestrial planets at ${\rm[Fe/H]}=0$, i.e. $f_{\rm TP}=0.14$ \citep{Boley+24}.
Note that $f_{\rm TP}=0$ below each minimum ${\rm[Fe/H]}$ for models F1-F3, while in E1 $f_{\rm TP}$ is extrapolated down to ${\rm[Fe/H]}=-2.5$, and in E2 $f_{\rm TP}<10^{-5}$ for ${\rm[Fe/H]}\sim-0.8$. 

\subsection{Galactic sources of sterilization}\label{sec:life-threats}
In principle, the emission from a nearby SN or GRB could induce ozone depletion in the Earth's atmosphere, so that life on the surface would be exposed to deadly doses of solar UV light, and possibly go extinct \citep[e.g.,][]{Ruderman74,Thorsett95}.
Here, we compute the volume sterilized by a single source as
\begin{equation}
    V_{\rm ster}=\dfrac{4\pi}{3}d_{\rm cr}^{3}\pqty{1-\cos\dfrac{\theta}{2}},
    \label{eq:Vster}
\end{equation}
where $d_{\rm cr}$ is the maximum distance up to which the source emission could destroy life on a planet, and $\theta$ takes into account for the geometry of the emission.
In the case of Type II SNe (SNe II) we adopt $d_{\rm cr}=14\parsec$, accounting for both standard~\citep{Gehrels+03,MelottThomas11,Fields+20} and with a strong X emission\footnote{These only represent $7\%$ of the whole SN II population~\citep{Li+11,ColdHjort23}.}~\citep{Brunton+23} SNe II, while for Type Ia SNe (SNe Ia) we estimate $d_{\rm cr}\approx100\parsec$ based on observational data from \citet{Churazov+15} (for details see the Appendix \ref{appx:dIa}).
We set $\theta=\pi$ for all SNe (spherical symmetry).
Finally, for GRBs we adopt $d_{\rm cr}=1\kilopc$ \citep{Thomas+05a,Thomas+05b} and $\theta=10\degree$ in order to take into account their typical collimation~\citep[see][]{Goldstein+16}.

\subsection{Sterilization-free environments}\label{sec:ster-free}
In this section we show how we compute $P_{\rm surv}$, defined as the probability of terrestrial planets in dSphs to survive galactic threats long enough to allow possible biological development.
First, we convert the cumulative SFH into a star formation rate (SFR) as a function of time, adopting a mass-to-light ratio $M/L=1$ and a time step\footnote{Assuming different time steps does not affect our findings.} of $100\megayr$.
Assuming that stars are homogeneously distributed within the volume of the dSphs, we compute the stellar mass fraction that forms at a time $t$ and that is unaffected by sterilizing effects of galactic sources (see Sec.~\ref{sec:life-threats}) between $t$ and $t+\deltat$:
\begin{equation}
    f_{\rm surv}\pqty{t,\deltat}=\dfrac{M_*\pqty{t}}{M_*^{\rm tot}}\pqty{1-\dfrac{V_{\rm fill}\pqty{t,\deltat}}{V_{\rm gal}}},
    \label{eq:fsurv}
\end{equation}
where $M_*\pqty{t}$ is the stellar mass formed at $t$, $M_*^{\rm tot}$ is the total stellar mass observed today, and $V_{\rm gal}=\pqty{4\pi/3}R_{\rm gal}^{3}$ is the volume of the galaxy with $R_{\rm gal}=2R_{1/2}$. 
The quantity $V_{\rm fill}\pqty{t,\deltat}$ is the dSph volume affected by lethal emission between $t$ and $t+\deltat$, and is computed as
\begin{equation}
    V_{\rm fill}\pqty{t,\deltat}=\sum_{j}V_{\rm ster,\it j}\int_{t}^{t+\deltat}R_{j}\pqty{t'}dt',
    \label{eq:Vfill}
\end{equation}
where $V_{\rm ster, \it j}$ is the volume sterilized by the {\it j}th source (SN II, SN Ia, or GRB) from \cref{eq:Vster}, and  $R_j\pqty{t'}$ is the rate of the {\it j}th source at time $t'$ computed using analytic prescriptions (see Appendix \ref{appx:rates} for details).
If $V_{\rm fill}\pqty{t,\deltat}>V_{\rm gal}$ all stars (and planets) that form at $t$ are bathed by lethal emission between $t$ and $t+\deltat$, then $f_{\rm surv}\pqty{t,\deltat}\equiv0$.
Note that \cref{eq:Vfill} overestimates $V_{\rm fill}\pqty{t,\deltat}$, as the regions sterilized by different sources could partially overlap. Consequently, \cref{eq:fsurv} might underestimate $f_{\rm surv}\pqty{t,\deltat}$.

We integrate \cref{eq:Vfill} for both the timescales typically assumed for the appearance of life on Earth in its embryonic form ($\deltat=1\gigayr$) and its evolution into an intelligent one ($\deltat=4\gigayr$).

Finally, by integrating \cref{eq:fsurv} between the time at which the dSph galaxy begins forming stars ($t_{\rm form}$) and the present day ($t_{\rm today}$), we get the global mass fraction of stars that are not irradiated from any galactic source at a distance $d<d_{\rm cr}$ for a time $\deltat$. 
We interpret this as $P_{\rm surv}$, i.e. the probability of putative terrestrial planets in the dSph to survive sterilizing effects from nearby galactic sources for a time $\deltat$.
\begin{equation}
    P_{\rm surv}=\int_{t_{\rm form}}^{t_{\rm today}}f_{\rm surv}\pqty{t',\deltat}dt'.
    \label{eq:Psurv}
\end{equation}
Note that, for a given SFH, the value of $P_{\rm surv}$ changes when adopting a different $\deltat$.

\section{Results} \label{sec:results}

In this section, we present the results of our model, starting with those reported in the right panel of Fig.~\ref{fig:sfh+mdf}.
Here we show, for dSph galaxies with different luminosities, the cumulative mass fraction of stars that survive life-harmful events for a time $\deltat$, obtained by integrating~\cref{eq:fsurv}.
The area below each curve is highlighted in green.
First, we see that these cumulative fractions decrease with the luminosity of the dSphs. 
In particular, in all UFDs the majority of stars contribute to the cumulative $f_{\rm surv}$, which is already $>80\%$ of the total around $9\gigayr$ ago and independent of the assumed $\deltat$.
In more luminous classical dSphs, this fraction is initially zero, it remains $\leq 60\%$ for dSphs with intermediate luminosities (if $\deltat=1$~Gyr), and it reaches the minimum for the brightest dSph galaxy, Fornax.

These results can be explained as follows: the higher the luminosity of a dSph (and so the higher its final stellar mass) the higher its SFR\footnote{$\textrm{SFR}<10^{-4}\Msun\yr^{-1}$ in UFDs while $\textrm{SFR}\approx (10^{-4}-10^{-3})\Msun\yr^{-1}$ in classical dSphs; see also \cite{Salvadori+14}.} and in turn the rate of disruptive events, resulting in a larger sterilized volume fraction (see \cref{eq:Vfill}, \cref{eq:SNR}, and \cref{eq:SNRIa}).

Note that in dSphs with intermediate luminosity (Dra, UMi, and Car) the choice of $\deltat$ can strongly affect the fraction of stars that contribute to $f_{\rm surv}$. The effect is particularly evident in Carina, which has a peculiar SFH made up of two main bursts of star formation, the last and most intense one occurring $<4$~ Gyrs ago.
We should also note that, despite the different luminosities, all UFDs show similar (cumulative) values of $f_{\rm surv}$.
This is due to their peculiar SFH. Since almost all stars in UFDs formed in the first billion years of cosmic evolution, the rate of SNe and GRBs in these galaxies rapidly drops $\sim30\megayr$ after\footnote{The lifetime of $8\Msun$ stars is $\approx25\megayr$ \citep{Raiteri+96}.} the quenching of star formation. Furthermore, the contribution of SNe Ia is negligible because of the low SFR in UFDs. 
Ultimately, the total number of destructive events in UFDs is limited in size and time. 
Thus, only a small portion of the galaxy is sterilized and more than $80\%$ of the stars that formed in the UFD during the first billion years of cosmic evolution are hosted in a galactic environment free from destructive events, up to the present day.

\subsection{Terrestrial planets in LG dSphs}\label{sec:Pform}
\begin{figure}
    \centering
    \includegraphics[width=\linewidth]{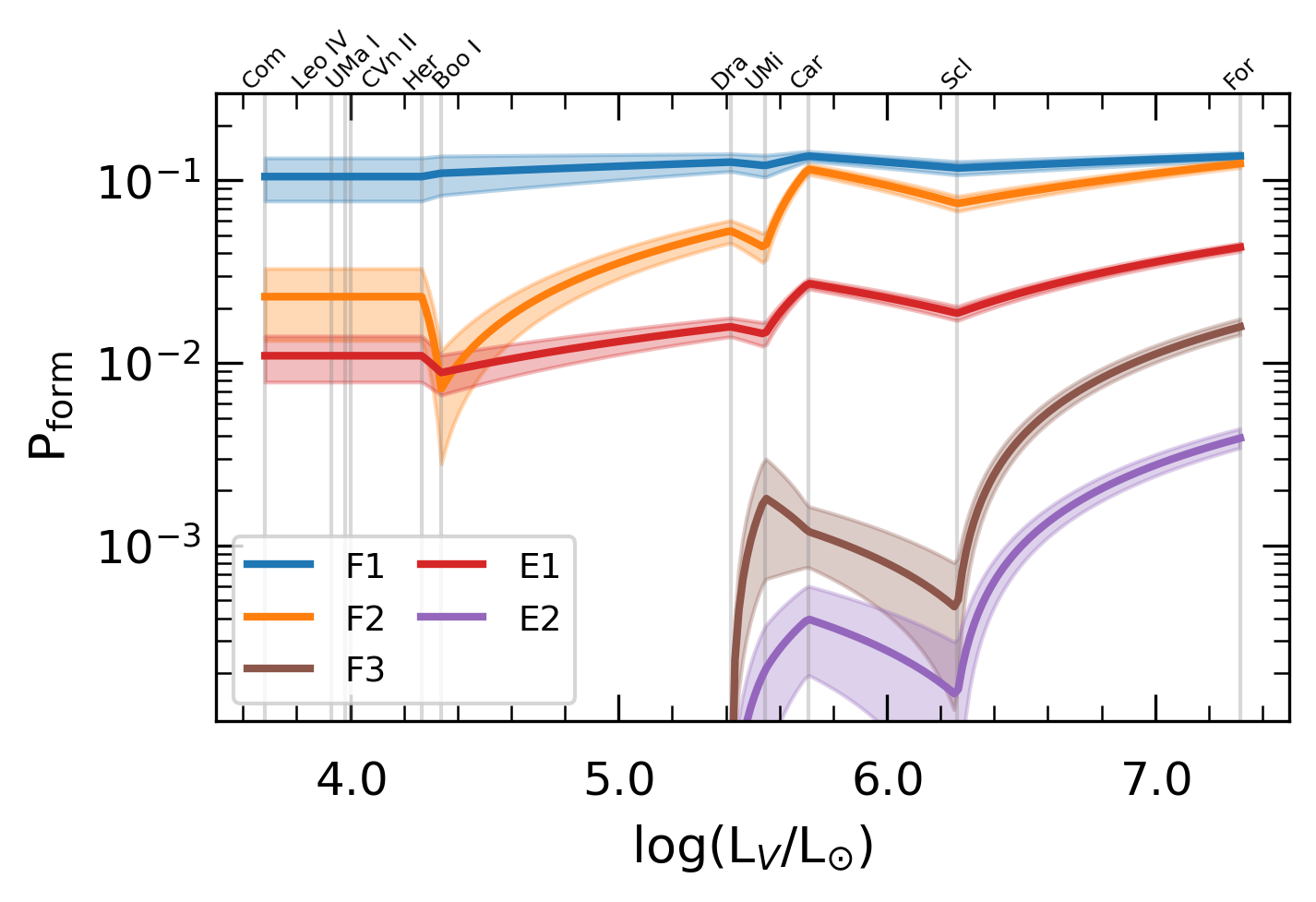}
    \caption{Probability of terrestrial planets formation in local dSphs, $P_{\rm form}$, as a function of the dSph luminosity. Each curve corresponds to a specific model for $f_{\rm TP}$ (see labels and Fig.~\ref{fig:eff_pl}), and is the linear interpolation of values computed for dSphs in our sample (identified by vertical lines). Shaded areas represent the uncertainties.}
    \label{fig:Pform}
\end{figure}

In Fig.~\ref{fig:Pform} we show the probability of terrestrial planets formation around stars in local dSphs, $P_{\rm form}$, as a function of the dSph's luminosity, $L_V$. The five curves show the results obtained by assuming different $f_{\rm TP}\pqty{\rm[Fe/H]}$ (see Sec.~\ref{sec:plformation}).
Each curve is the linear interpolation between the values of $P_{\rm form}$ computed for each dSph in our sample.
Shaded areas show Poissonian uncertainties, which are lower for higher $L_{V}$ because of the larger number of available measurements.

In Fig.~\ref{fig:Pform} we see that the results can be divided into two groups showing different trends.
Models that allow the formation of terrestrial planets around stars with ${\rm[Fe/H]}\lsim-1$ and assume a constant (F1, F2) or an increasing (E1) $f_{TP}$ with ${\rm[Fe/H]}$, yield $P_{\rm form}\approx1\%-10\%$ for all dSphs in our sample.
In these cases $P_{\rm form}$ is maximum ($\geq 10\%$) and (almost) independent of $L_V$ for model F1, which assumes $f_{\rm TP}=0.14$ for ${\rm[Fe/H]}\geq-2.5$, while it increases with $L_V$ and is $\leq10\%$ for both models F2 and E1, which assume different shapes for $f_{\rm TP}$ (see Fig.~\ref{fig:eff_pl} and Sec.~\ref{sec:plformation}).
In contrast, if terrestrial planets can only form around stars with ${\rm[Fe/H]}\geq-0.6$ (models F3 and E2), then they should not exist in UFDs. This is due to the fact that UFDs are the most metal-poor galaxies, lacking stars at these iron abundance values as shown in the left panels of Fig.~\ref{fig:sfh+mdf}.
Note that these models predict $P_{\rm form}<1\%$ for classical dSphs, with large variations for $L_V\gtrsim10^{5.4}\Lsun$ and the maximum $P_{\rm form}$ reached for the most luminous For dSphs. This is because the MDFs of classical dSphs have a peak increasing with $L_V$, but they are all poorly populated at ${\rm[Fe/H]}\geq-0.6$. 

Ultimately, two major scenarios emerge for the formation of terrestrial planets in dSphs: (i) they form in both UFDs and classical dSphs with a probability of $1\%-10\%$ if $f_{\rm TP}>0$ for ${\rm[Fe/H]}\lsim-1$; (ii) they only form in luminous classical dSphs if $f_{\rm TP}>0$ for ${\rm[Fe/H]}\geq-0.6$ with a lower probability $P_{\rm form}=0.01\%-1\%$.

\subsection{Life-harmful events in LG dSphs}\label{sec:Psurv}
\begin{figure}
    \centering
    \includegraphics[width=\linewidth]{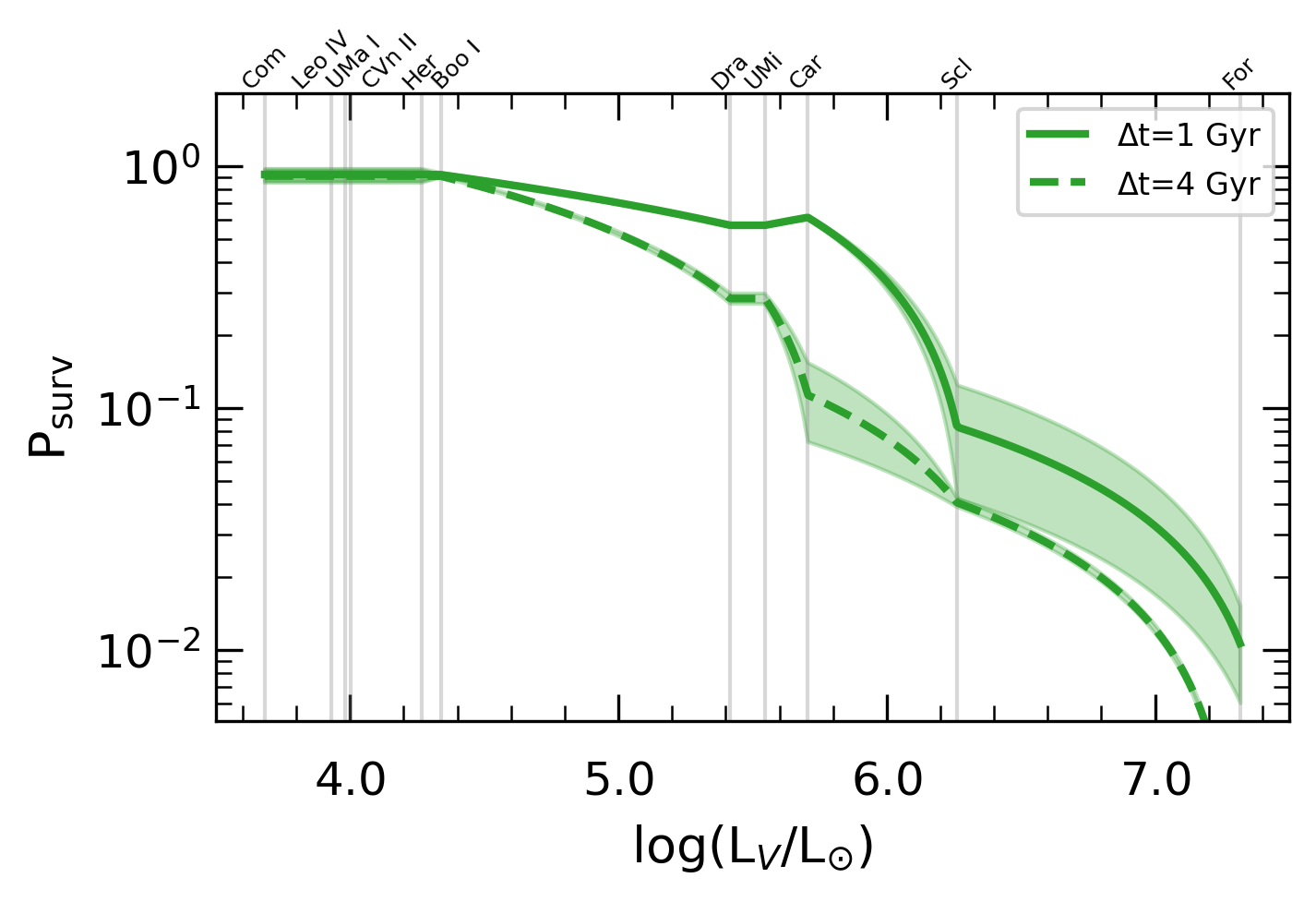}
    \caption{Probability of putative planets in local dSphs to survive, for a time $\deltat$, against destructive galactic events, $P_{\rm surv}$, as a function of the dSph luminosity. 
    Curves are obtained as the linear interpolation of the values computed for dSphs in our sample (identified by vertical lines), in the cases of $\deltat=1\gigayr$ (solid) and $\deltat=4\gigayr$ (dashed). Shaded areas represent the uncertainties.}
    \label{fig:Psurv}
\end{figure}

In Fig.~\ref{fig:Psurv} we show the probability of newly formed planets in dSphs to survive life-harmful events, $P_{\rm surv}$, as a function of $L_V$. The two curves correspond to different timescales for possible biological development. 
Due to similar but scattered values of $P_{\rm surv}$ in UFDs, here we show the average value for all of them, with the only exception of Bo\"{o}tes~I.
Overall, we see that $P_{\rm surv}$ is roughly constant ($P_{\rm surv}\sim1$) for $L_V<10^{5}\Lsun$ (i.e. for UFDs), and then it decreases for increasing $L_V$ down to $P_{\rm surv}\sim0.01$ (or $P_{\rm surv}=0$ for $\deltat=4\gigayr$) for the Fornax dSph. 
This decreasing trend can be explained as follows. More luminous dSphs have larger SFRs, yielding higher rates of destructive events; hence, despite being larger than UFDs, they are more significantly affected by life-harmful events, resulting in a lower $P_{\rm surv}$ (see \cref{eq:fsurv,eq:Psurv}).
\begin{figure*}
    \centering
    \includegraphics[width=\linewidth]{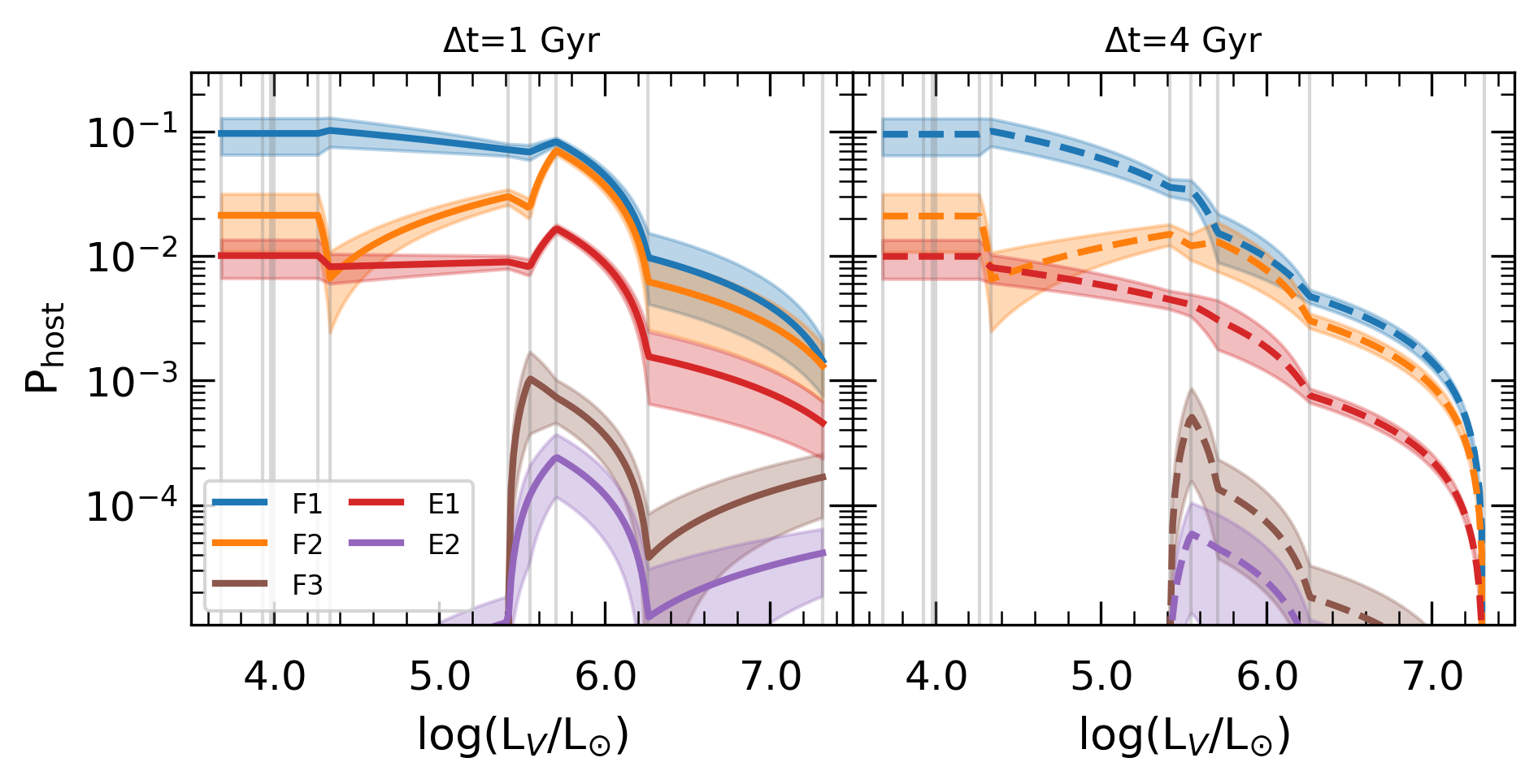}
    \caption{Probability of LG dSphs to host terrestrial planets dwelling for $\deltat$ in a safe environment, $P_{\rm host}$, as a function of the dSph luminosity. Curves are obtained through \cref{eq:Phost}, combining $P_{\rm form}$ resulting from different models of $f_{\rm TP}$ (see Fig.~\ref{fig:Pform}), and $P_{\rm surv}$ for $\deltat=1\gigayr$ (left panel, solid) and $\deltat=4\gigayr$ (right panel, dashed). Shaded areas represent the uncertainties from the error propagation in \cref{eq:Phost}.}
    \label{fig:Phost}
\end{figure*}

We see that when we consider a longer $\deltat$, we get lower values of $P_{\rm surv}$ for $L_V\gtrsim10^{4.3}\Lsun$.
Indeed, with respect to UFDs, more luminous dSphs have more stars, but they are formed on much longer timescales (see the right panels of Fig.~\ref{fig:sfh+mdf}). As a result, the rate of SNe and GRBs is higher and has a much broader distribution in cosmic time.

\subsection{Safe terrestrial planets in LG dSphs}\label{sec:Phost}

Fig.~\ref{fig:Phost} summarizes the final results of our study, showing the probability of dSphs with different $L_V$ to host safe terrestrial planets, 
$P_{\rm host}$, for all the different models considered in our study (different
$f_{\rm TP}\pqty{\rm[Fe/H]}$ and $\deltat$).
Let us start with the left panel, which illustrates the results obtained for $\deltat=1\gigayr$, i.e. the timescale for the appearance of life on Earth in its embryonic form.
We note once again that the results can be divided into two major groups.
In models where planet formation is possible for ${\rm[Fe/H]}\lsim-1$ (F1, F2, and E1) $P_{\rm host}\approx0.8\%-10\%$ up to $L_V\simeq10^{5.7}\Lsun$, with small variations for each model.
For higher $L_V$ it decreases down to $P_{\rm host}\approx0.04\%-0.1\%$ in the Fornax dSph.
Conversely, if planets form only for ${\rm[Fe/H]}\geq-0.6$ (models F3 and E2) $P_{\rm host}=0$ for $L_V<10^{5}\Lsun$ (hence in all in UFDs), while $P_{\rm host}\approx0.001\%-0.1\%$ in more luminous dSphs.
When we consider $\deltat=4\gigayr$, we obtain similar trends with a few notable differences. While $P_{\rm host}$ is independent of $\deltat$ for $L_V\lsim10^{4.5}\Lsun$, we note that for higher luminosities it assumes lower values with respect to the $\deltat=1\gigayr$ case. In particular, for dSphs with intermediate $L_V$ (Dra, UMi, and Car) when we assume $\deltat=4\gigayr$ $P_{\rm host}$ is a factor of $\sim5$ lower, while in the most luminous dSph, Fornax, $P_{\rm host}=0$.
Interestingly, according to our model a sort of ``luminosity sweet spot" exists for $10^{5.4}\Lsun\leq L_V\leq10^{6.6}\Lsun$ where $P_{\rm host}>0$, independently of which $f_{\rm TP}$ or $\deltat$ we assume. 

Ultimately, our model predicts two possible major scenarios, which are mainly driven by $P_{\rm form}$:
(i) if $f_{\rm TP}>0$ for ${\rm[Fe/H]}\leq-1$, both UFDs and classical dSphs are able to safely host terrestrial planets long enough for possible biological development, with a probability $P_{\rm host}\approx0.1\%-10\%$; 
(ii) if $f_{\rm TP}>0$ for ${\rm[Fe/H]}\geq-0.6$, terrestrial planets with possible biological development can only form in classical dSphs and with a much lower probability, $P_{\rm host}\approx0.001\%-0.1\%$.
\\
\\
\section{Discussion}\label{sec:discussion}

In our work we explored the formation of terrestrial planets in LG dSphs, and their long-term galactic exposure to destructive events.
Indeed, such planets would represent suitable places in a galaxy for the possible appearance and development of life.
However, whether these planets might be actually habitable would also depend on planetary-scale physical aspects, such as the host star's spectral type, planetary orbital distances, and the stellar habitable zone, whose modeling is beyond the scope of this work.
Nevertheless, we qualitatively discuss how their interplay might affect our results.
Furthermore, we discuss on the role of refractory elements (Mg and Si, other than just Fe) in planet formation in metal-poor environments, and consider the possible investigation of other ancient LG environments.

\subsection{Stellar spectral type and habitable zone}
The results we used to model $f_{\rm TP}$ are valid only for close-in planets ($\approx0.01-0.5\au$) orbiting FGK stars (see Sec.~\ref{sec:plformation} and references therein).
However, the majority of stars in our sample are K and M dwarfs. 
Indeed, stars with $m\leq0.8\Msun$ belong to the K and M spectral classes \citep[e.g.][]{Baraffe+15}, and have lifetimes $>9\gigayr$ \citep{Raiteri+96}.
From the SFHs in Fig.~\ref{fig:sfh+mdf} we find that all stars in UFDs formed more than $9\gigayr$ ago, and hence they are K or M. The same argument can be applied to $\geq 95\%$ of the stars in Draco, Ursa Minor, and Sculptor, and to $\geq 40\%$ of the stars in Carina and Fornax. 
Thus, in this work, we extend all our models of $f_{\rm TP}$ to M dwarfs and we make a conservative assumption by maintaining the same normalization, despite in the MW the observed planet occurrence rate being higher for M dwarfs \citep{DressingCharbonneau15,Gore+24} than for FGK stars \citep{Zink+23,Boley+24}.
Furthermore, planet formation in low-metallicity environments is expected to occur relatively close to the host star ($<0.5\au$), as the disk mass is limited \citep[e.g.][]{JohnsonLi12}.
Interestingly, for such close-in planets, it would be possible to orbit within the stellar habitable zone of K and M dwarfs, which for an Earth-like planet includes orbits from $\approx0.03\au$ up to $\lsim1\au$ \citep[e.g.][]{Kopparapu+13}.
Ultimately, it would be possible for putative terrestrial planets around K and M dwarfs in local dSph galaxies to orbit within the stellar habitable zone. 
If so, according to our model these potentially habitable planets would have good chances to evolve in a safe environment from major galactic events, providing a favorable place for life to appear and evolve.
To quantify the probability for life to appear we should model planet formation within the stellar habitable zone as a function of the host star's mass and metallicity. This is beyond the scope of this study, but we plan to do it in a dedicated future work.

\subsection{Milky Way bulge and Globular Clusters}
In the LG there are other ancient environments that could be explored in terms of planet formation and long-time exposure to galactic threats, such as the MW bulge and globular clusters (GCs).
Despite being more metal-rich \citep[e.g.,][]{Barbuy+15}, the MW bulge is believed to be inhospitable since early cosmic times because of the high rates of (i) lethal events \citep{Spinelli+21}, (ii) ionizing photons emitted by the active galactic nucleus at the center of the MW \citep{Ambrifi+22}, and (iii) star-formation at early cosmic epochs driven by merging events \citep{Pagnini+23}.
On the other hand, GCs experience few (or single) bursts of star formation (like UFDs), but their stellar populations have higher ${\rm[Fe/H]}$ \citep[e.g.,][]{Forbes+10,Pancino+17,Belokurov+24}. 
Furthermore, GCs are $(1-2)\gigayr$ younger than UFDs \citep[e.g.,][]{Garro+24} and are expected to host high fractions of X-ray binaries \citep[e.g.,][]{Bellazzini+95}, which could represent an additional threat to life, and being dense environments their putative planetary systems might be disrupted by stellar encounters \citep[e.g.,][]{Sigurdsson92}.
Ultimately, investigating GCs as possible sites to safely host terrestrial planets represents an interesting application of our model, although their SFHs suggest that, likely, they would not be the first. 

\begin{figure}
    \centering
    \includegraphics[width=\linewidth]{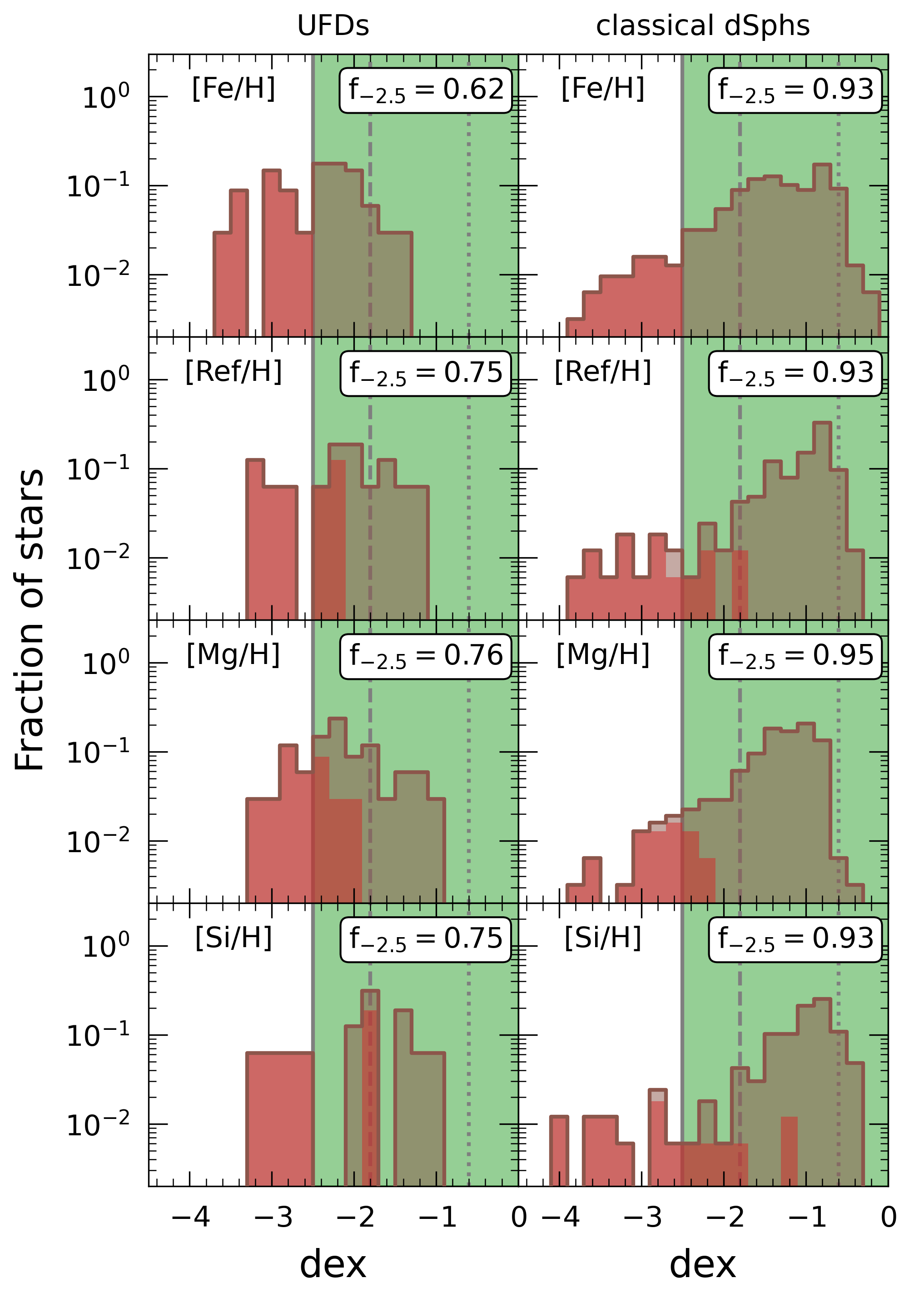}
    \caption{Normalized fraction of stars in UFDs (left) and dSphs (right) with different chemical abundances (see labels) obtained using the high-resolution measurements ($R>10^{4}$) in the SAGA database. All measurements are corrected for 1D NLTE effects, using \texttt{NLiTE} \citep{Koutsouridou25}.
    Red histograms show stars with ${\rm[Fe/H]}<-2.5$. In each panel we report the fraction of stars above $-2.5$ dex, $f_{-2.5}$ (see labels).
    Vertical lines and green shaded areas are the same as those in the left panels of Fig.~\ref{fig:sfh+mdf}.}
    \label{fig:zeff_mdf}
\end{figure}

\subsection{Refractory elements}

From our study it follows that UFDs might host terrestrial planets only if $f_{\rm TP}>0$ for ${\rm[Fe/H]}\lsim-1$ (see Fig.~\ref{fig:Pform}).
In this regard, we should note that at low ${\rm[Fe/H]}$ the iron abundance might not be a very good proxy for metallicity.
This becomes relevant if we consider that the seeds for planet formation are dust grains \citep[e.g.,][]{Adibekyan19,Testi+22}, and that their cores are made of magnesium and silicon, other than iron \citep[e.g.,][]{Kozasa+89,TodiniFerrara01}. 
Indeed, \citet{Gonzalez09} proposed using the refractory index ${\rm[Ref/H]}$, which account for Mg, Si, and Fe, to describe the planet formation probability as a function of metallicity \citep{Adibekyan+12a,Adibekyan+12b}.
How does the MDF of UFDs and classical dSphs change when we use ${\rm[Ref/H]}$ instead of ${\rm[Fe/H]}$? And how does it affect our scenarios for planet formation in local dSphs?
We selected stars in our sample with ${\rm[Mg/H]}$, ${\rm[Si/H]}$, and ${\rm[Fe/H]}$ high-resolution measurements. Then, we built distributions with respect to each element and ${\rm[Ref/H]}$. 
The results are shown in Fig.~\ref{fig:zeff_mdf}, where we also report the minimum ${\rm[Fe/H]}$ for which $f_{\rm TP}>0$ in each model.
In each distribution, the red histograms count stars with ${\rm[Fe/H]}<-2.5$.

We see that in UFDs $\approx35\%$ of stars with ${\rm[Fe/H]}<-2.5$ have ${\rm[Ref/H]}>-2.5$ because of an enhancement in ${\rm[Mg/H]}$ and/or ${\rm[Si/H]}$ (while in classical dSphs the increase is less than $1\%$).
Thus, in metal-poor environments such as UFDs, ${\rm[Fe/H]}$ is not a good proxy for metallicity, and it might lead to an underestimate of the number of planets in metal-poor galaxies.
Interestingly, UFDs are known to have the highest fraction of C-enhanced stars \citep{Rossi+23,Lucchesi+24}, which are polluted by the first supernovae and thus are also enhanced in Mg and Si \citep[see lower panels of Fig.~\ref{fig:zeff_mdf};][]{Vanni+23,Saccardi+23}. 
Moreover, recent simulations show that these first SN explosions could have produced abundant amount of water in their remnants \citep{Whalen+25}, also inducing the collapse of dense cores where several Earth masses of planetesimals are formed $\approx200\megayr$ after the Big Bang \citep{Vorobyov+25}. 
\\
\section{Summary and Conclusion}\label{sec:conclusions}
In this work we explored for the first time whether and when LG dSphs might have hosted terrestrial planets dwelling in a safe galactic environment during their subsequent long-term evolution.
To this end, we developed a novel and general semiempirical model that exploits the observed MDFs and SFHs to quantify for each galaxy the probability of forming terrestrial planets around its stars and for putative planets the probability of avoiding life-harmful irradiation from SNe and GRBs, during the expected timescale for life appearance ($\deltat=1\gigayr$) and evolution ($\deltat=4\gigayr$) on Earth. 
The model can be applied to every stellar system for which these quantities are measured.

From our model two major scenarios emerge: (i) In models where planet formation is possible for ${\rm[Fe/H]}\lsim-1$ (models F1, F2, and E1), both UFDs and classical dSphs safely host terrestrial planets long enough to possibly allow biological development, with a probability $P_{\rm host}\approx0.1\%-10\%$.
(ii) Conversely, if planets form only for ${\rm[Fe/H]}\geq-0.6$ (models F3 and E2), this should be possible only in classical dSphs with $L_V\geq10^{5.4}\Lsun$, where $P_{\rm host}\approx0.001\%-0.1\%$. In this case, planets should not exist in UFDs.

Noticeably, in scenario (i) UFDs would be the first local galaxies able to host terrestrial planets in a long-term safe galactic environment, as they form most of their stars during the first billion years of cosmic time, i.e. several billions of years before the formation of our own solar system.
In this sense, these ancient and small galaxies we observe today as relics of the early Universe might have become primordial life nurseries.

Skeptics may wonder about the assumptions made in our study. We showed that the choice of a $\deltat$ compatible with the timescale of the appearance (1~Gyr) or the evolution (4~Gyr) of life on Earth does not affect the general trends for $P_{\rm host}$, with the two exceptions of the Carina and Fornax dSphs.  
Furthermore, we checked that the values of $P_{\rm host}$ lie within the uncertainties if we assume a $10\x$ smaller volume for each dSph, i.e. if we mimic an inhomogeneous and more concentrated distribution of stars (see Appendix~\ref{appx:rgal}).

In conclusion, our work suggests that the minimum metallicity that triggers planet formation around a star and the probability distribution to form planets at different stellar ${\rm[Fe/H]}$, are key to understanding whether and when this could have happened for the very first time in the history of the LG. 
Any future search for planets around metal-poor stars (in the MW, but hopefully also in local dSphs) would help us to discriminate between the two scenarios predicted in this work, possibly shedding light on the most ancient systems that could have provided the minimum conditions to potentially host life. \\

\section*{Acknowledgements}
The authors want to thank the anonymous referee and the editor for their very constructive comments that have certainly improved this manuscript.
S.C. and S.S. acknowledge support by the European Research Council (ERC) Starting Grant NEFERTITI H2020/804240 (PI: Salvadori). L.T. is partly supported by the Italian Ministero dell’Istruzione, Università e Ricerca through the grant Progetti Premiali 2012-iALMA (CUP C52I13000140001), the European Union’s Horizon 2020 research and innovation program under the Marie Skłodowska-Curie grant agreement No. 823823 (DUSTBUSTERS), and the ERC via the ERC Synergy Grant ECOGAL (grant 855130).



\appendix

\section{The Rate of SNe II, SNe Ia and GRBs}\label{appx:rates}
Here we describe the analytical relations used to compute the rates of disruptive events in \cref{eq:Vfill}. 
The rate of SNe II is
\begin{equation}
    R_{\rm II}\pqty{t}=\nu\int_{\mstar_{\rm min}}^{\mstar_{\rm max}}\textrm{SFR}\pqty{t-\tau\pqty{\mstar}}\phi\pqty{\mstar}d\mstar
    \label{eq:SNR}
\end{equation}
where $\phi\pqty{\mstar}$ is the initial mass function (IMF), $\nu$ is the number of stars per unit stellar mass formed, and $\tau\pqty{\mstar}$ is the lifetime of stars with mass $\mstar$~\citep{Raiteri+96}.
We assume $\mstar_{\rm min}=8\Msun$ and $\mstar_{\rm max}=40\Msun$  - since more massive stars are expected to directly collapse into black holes preventing SN explosion~\citep{WoosleyWeaver95,LimongiChieffi18} - and a Larson-type IMF
\begin{equation}
    \phi\pqty{\mstar}\equiv\dfrac{dN}{d\mstar}\propto\mstar^{-1+\alpha}\exp\pqty{-\dfrac{m_{\rm ch}}{\mstar}},
    \label{eq:IMF}
\end{equation}
in the range $\bqty{0.1-100}\Msun$ with $\alpha=-1.35$ and $m_{\rm ch}=0.35\Msun$~\citep{Larson98}.

For the rate of GRBs, $R_{\rm GRB}\pqty{t}$, we use the same equation~(\ref{eq:SNR}) but with $\mstar_{\rm max}=100\Msun$, and multiply the final value for a factor $f_{\rm GRB}=0.1$. 
This accounts for the unknown fraction of massive stars that can originate a GRB either because (i) they are rapidly rotating~\citep[long GRBs, e.g., ][]{MacFadyenWoosley99} or because (ii) they pertain to a binary system in which both companions evolve as compact objects that will later merge~\citep[short GRBs, e.g.,][]{Berger14}.

For the SN Ia rate we use the relation from ~\citet{Matteucci+06}:
\begin{equation}
    R_{\rm Ia}\pqty{t}=\nu\int_{\tau_{\rm min}}^{\min\pqty{t,\tau_{\rm max}}}A_{\rm Ia}\textrm{SFR}\pqty{t-\tau}{\textrm{DTD}}\pqty{\tau}d\tau,
    \label{eq:SNRIa}
\end{equation}
where $\textrm{DTD}\pqty{\tau}$ is the delay-time distribution function from~\citet{Mannucci+06}, which accounts for the time needed for SN Ia to occur after star formation. Following~\citet{Matteucci+06} we assume that the minimum (maximum) delay time for the occurrence of an SN Ia, $\tau_{\rm min}$ ($\tau_{\rm max}$), is equal to the lifetime of a $8\Msun$ ($0.8\Msun$) star.
The quantity $A_{\rm Ia}$ is the fraction of binary systems that produce SN Ia. 
In consideration of the lack of constraints for the observed rate of SNe Ia in dSph galaxies, we assume $A_{\rm Ia}=2.5\x10^{-3}$ to reproduce the observed rate of $\approx\pqty{0.3/100}\yr^{-1}$ in the MW~\citep{Cappellaro+99}.
\begin{figure}
    \centering
    \includegraphics[width=\linewidth]{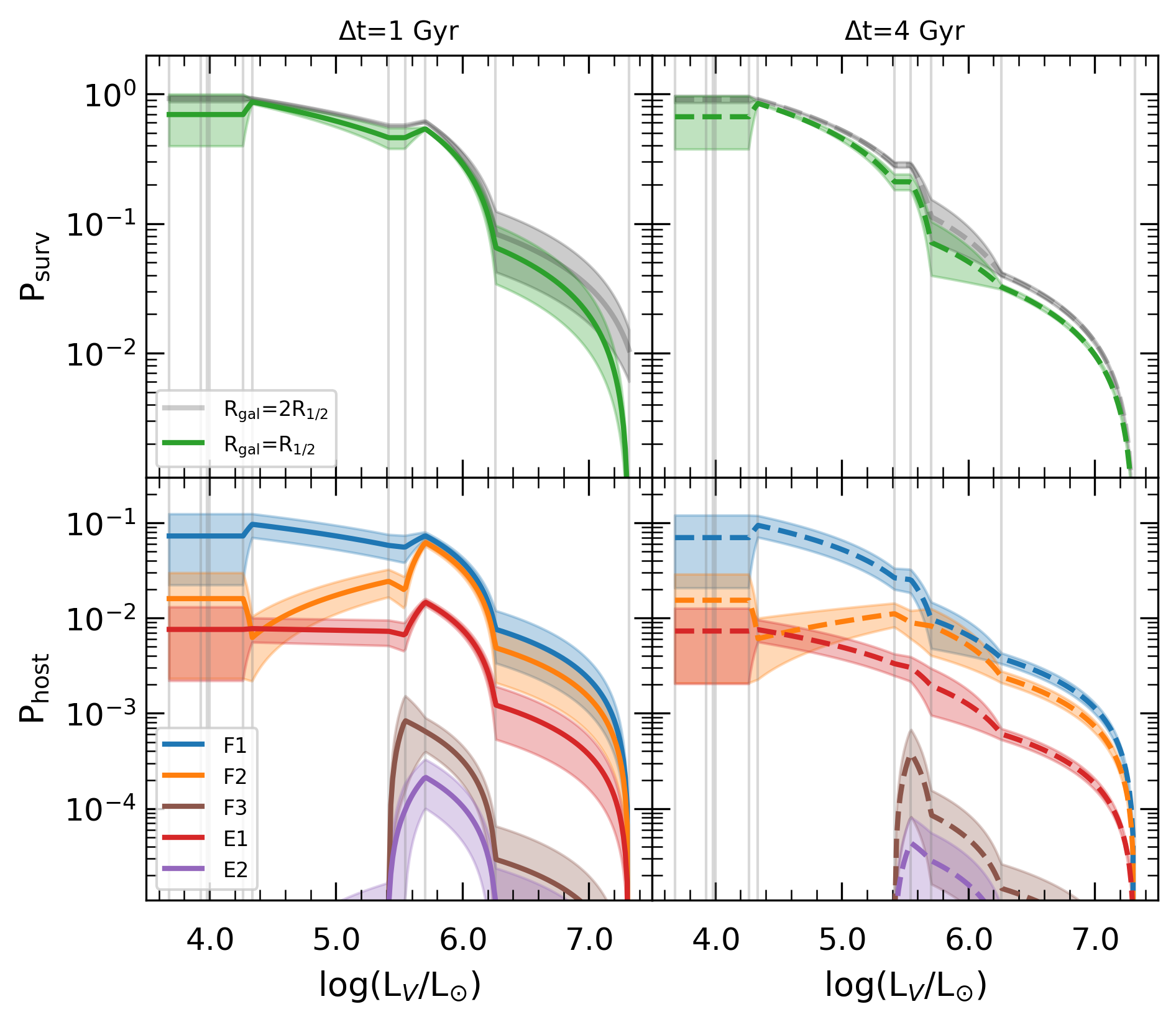}
    \caption{Top: comparison between $P_{\rm surv}$ as a function of $L_V$, computed for a galaxy radius $R_{\rm gal}=R_{1/2}$ (green) and $R_{\rm gal}=2R_{1/2}$ (gray, same as Fig.~\ref{fig:Psurv}).
    Bottom: the curves are the same as those in Fig.~\ref{fig:Phost}, but obtained using $R_{\rm gal}=R_{1/2}$ for the calculation of $P_{\rm surv}$ (green curves in top panels).
    Left panels are for $\deltat=1\gigayr$ (solid curves) and right panels are for $\deltat=4\gigayr$ (dashed).}
    \label{fig:appx_rh}
\end{figure}

\section{Critical distance for type Ia SNe}\label{appx:dIa}
Type Ia SNe emit in the gamma and in principle could induce biological damage on a planet if the time-integrated flux is ${\cal F}_{\gamma}\geq{\cal F}_{\gamma}^{\rm cr}$, where ${\cal F}_{\gamma}^{\rm cr}=10^{8}\erg{\rm cm}^{-2}$ is the estimated critical value above which there might be lethal consequences for life on Earth~\citep{Thomas+05a,Thomas+05b,Ejzak+07}.
We exploit available observational data of the SN Ia SN2014J (${\cal F}_{\gamma}\approx7\x10^{-2}\erg{\rm cm}^{-2}$, $d=3.5\megapc$,~\citealp{Churazov+15}) to estimate the equivalent distance at which the time-integrated flux of SN2014J would equal the critical value (M. Richmond 2025, private communication), as follows
\begin{equation}
    d_{\rm cr,Ia}=\pqty{\dfrac{{\cal F}_{\gamma}}{{\cal F}_{\gamma}^{\rm crit}}}^{1/2}d\approx100\parsec.
\end{equation}

\section{Inhomogeneous stellar distribution}\label{appx:rgal}
Here we discuss how our results change if we assume a radius for the dSph galaxies that is $2$ times smaller: $R_{\rm gal}=R_{1/2}$. This mimics an inhomogeneous, but more concentrated, distribution of stars in the dSph galaxies. 

From the top panels of Fig.~\ref{fig:appx_rh} we see that if we assume $R_{\rm gal}=R_{1/2}$ (green) the results are roughly consistent with the case of $R_{\rm gal}=2R_{1/2}$ (gray), for both $\deltat=1\gigayr$ (in this case with the exception of the Fornax dSph) and $\deltat=4\gigayr$. 
Consequently, since the values of $P_{\rm form}$ have not changed in \cref{eq:Phost}, the same conclusion holds for $P_{\rm host}$ (bottom panels of Fig.~\ref{fig:appx_rh}).
We note that in UFDs $P_{\rm surv}$ decreases from $\approx0.90\pm0.06$ to $\approx0.70\pm0.30$; however this is solely driven by the two smallest UFDs (Com and CVn II), which have $R_{1/2}\approx70\parsec$ and thus can be totally sterilized if the dSph radius is halved.
For $L_V\gsim 10^{4.3}\Lsun$ the difference is $\lsim 15\%$. As a consequence, $P_{\rm host}$ decreases for all dSphs too, yet it remains consistent within the uncertainties. We showed that $P_{\rm host}$ lies within the uncertainties if we adopt a $2$ times smaller $R_{\rm gal}$.


\bibliography{mybib}{}
\bibliographystyle{aasjournal}


\listofchanges

\end{document}